\documentclass[aps,prb,twocolumn]{revtex4}
\usepackage{graphicx}
\usepackage{dcolumn}
\usepackage{amsmath}
\usepackage{amssymb}
\usepackage{float}
\usepackage{epstopdf}
\def\erfc{{\rm erfc}}
\def\erf{{\rm erf}}
\def\rv{{\bf r}}
\def\sv{{\bf s}}
\def\xv{{\bf x}}

\def\kv{{\bf k}}
\def\qv{{\bf q}}
\def\pv{{\bf p}}

\def\e{\epsilon}
\def\ec{\epsilon_c}
\def\ex{\epsilon_x}
\def\exc{\epsilon_{xc}}

\begin{document}
\title{A local density functional for the short-range part of the 
electron-electron interaction}
\author{Lorenzo Zecca, Paola Gori-Giorgi,\footnote{present address: 
Laboratoire de Chimie Th\'eorique,
Universit\'e Pierre et Marie Curie, Paris, France} Saverio Moroni, and Giovanni
B. Bachelet}
\affiliation{INFM Center for
  Statistical Mechanics and Complexity, and
Dipartimento di Fisica, Universit\`a di Roma ``La Sapienza,'' 
Piazzale A. Moro 2, 00185 Rome, Italy}
\date{\today}
\begin{abstract}
Motivated by recent suggestions --to split the electron-electron interaction
into a short-range part, to be treated within the density functional
theory, and a long-range part, to be handled by other techniques-- we compute, 
with a diffusion Monte Carlo method, the ground-state energy of a uniform
electron gas with a modified, short-range-only electron-electron
interaction $\erfc(\mu r)/r$, for different
values of the cutoff parameter $\mu$ and of the electron density. After 
deriving some exact limits, we propose an analytic representation of 
the correlation energy which accurately 
fits our Monte Carlo data and also includes, by construction, these
exact limits, thus providing a reliable ``short-range local-density 
functional''. 
\end{abstract}
\maketitle
\section{Introduction and summary of results}
Density functional theory\cite{kohnnobel,science,FNM}
 (DFT) is nowadays the most
widely used method for electronic structure calculations, 
in both condensed matter physics and quantum chemistry, thanks
to the combination of low computational cost and remarkable accuracy for a wide variety of chemical bonds and solid-state systems.
There are, however, notable exceptions to such an accuracy.
For example, even the best available approximations of the exchange-correlation functional, the key ingredient of the DFT, fail  to recover long-range van der Waals interactions,\cite{rydberg,dobson,perdewtao}
are not completely safe for the description of the hydrogen bond
\cite{fuchs2} and have intrinsic problems with situations of
near degeneracy (when two sets of orbitals happen to have very close energies).
\cite{Sav1,sav2002} More generally, the ``chemical accuracy'' (the accuracy needed to predict the rates of chemical reactions) has not yet been reached. For all these reasons  the
search for new approximate functionals, or even new ways of exploiting the basic ideas and advantages of the DFT, is very 
active.\cite{science,rydberg,dobson,perdewtao,fuchs2,sav2002} 
\par
In this context several
authors\cite{kohn,sav2002,Sav1,scuseria} 
have suggested to
split the electron-electron interaction into a short-range part,
to be treated within the DFT, and a long-range part, to be handled
by other techniques. The motivation behind these ``mixed schemes" is that the DFT, even in the simplest local-density approximation
(LDA), provides an accurate description of the
short-range electron-electron repulsion,\cite{ontop} while other techniques which give
a poor description of short-range properties, 
 like the configuration
interaction (CI) method or the random-phase approximation 
(RPA),\cite{fuchs,fuchs3}
can, instead, accurately capture long-range correlation effects.
\par
\begin{figure}
\includegraphics[width=6.6cm]{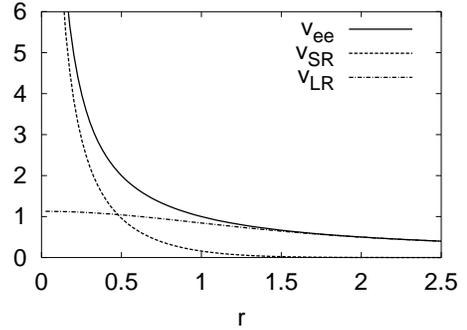} 
\caption{Splitting of the Coulomb electron-electron repulsion
$v_{ee}=1/r$ into a short-range (SR) part and a long-range (LR)
part, according to Eqs.~(\ref{eq_srpotential})-(\ref{eq_lrpotential}),
with $\mu=1$.}
\label{fig_erf}
\end{figure}

Of course there is no unique way to split the Coulomb potential
$1/r$ into a short-range (SR) and a long-range (LR) part.  The error function and its complement
\begin{eqnarray}
& & v_{ee}(r)=\frac{1}{r}=v_{\rm SR}(r)+v_{\rm LR}(r), \nonumber \\
& & v_{\rm SR}(r)=\frac{\erfc(\mu r)}{r}, 
\label{eq_srpotential}\\
& & v_{\rm LR}(r)=\frac{\erf(\mu r)}{r}, 
\label{eq_lrpotential}
\end{eqnarray}
 have been already used for this purpose \cite{sav2002,Sav1,scuseria} 
(see Fig.~\ref{fig_erf}), and we stick to this choice, 
which yields analytic matrix elements for both gaussians and plane waves, i.e., the most common basis functions in quantum chemistry and solid-state physics, respectively. This form still leaves room for some arbitrariness: the choice of the most convenient cutoff parameter $\mu$, which  may be different for different ``mixed schemes".
\par
The combination of a short-range DFT calculation and
a different treatment of the long-range part of the 
electron-electron interaction can be founded on a rigorous 
basis through the adiabatic connection
formalism.\cite{Sav1,sav2002,khon,adiabatic} Depending on the specific problem addressed
(van der Waals forces, near-degeneracy,...),
 and thus on the particular approach to the long-range part of the electron-electron 
 interaction, different ``mixed schemes" have been proposed.\cite{Sav1,sav2002,kohn} 
But in all of them, as in standard DFT, a crucial role is played by the
exchange-correlation functional, which now must be built for a modified 
electron-electron interaction.
The schemes of Refs.~\onlinecite{kohn,scuseria} need a
pure short-range functional, $E_{xc}^{\rm SR}[n]$, whose
LDA version is given by
\begin{equation}
E_{xc}^{\rm SR,\, LDA}[n]=\int n(\rv)\, \exc(n(\rv),\mu)\,d\rv,
\label{eq_LDASR}
\end{equation}
where $\exc(n,\mu)$ is the exchange-correlation energy per electron of
a uniform gas of density $n$ interacting with a short-range potential  
like Eq.~(\ref{eq_srpotential}). The value of $\mu$ 
in Eq.~(\ref{eq_LDASR}) can be either a constant, or, possibly,
a convenient function of the density, $\mu=\mu(n(\rv))$.\cite{savin_millanta}
The local functional $\exc(n,\mu)$ is the
quantity which we provide in this paper. We start from a jellium-like hamiltonian
(in Hartree atomic units used throughout this work)
\begin{equation}
H  =  -\frac{1}{2}\sum_{i=1}^N \nabla^2_{\rv_i}+V_{ee}^{\rm SR}
+V_{eb}^{\rm SR}+V_{bb}^{\rm SR},
\label{eq_ham}
\end{equation}
where $V_{ee}^{\rm SR}$ is the modified electron-electron interaction
\begin{equation}
V_{ee}^{\rm SR}  =  \frac{1}{2}\sum_{i\ne j=1}^N\frac{\erfc(\mu|\rv_i-\rv_j|)}
{|\rv_i-\rv_j|}, 
\label{eq_Vee}
\end{equation}
$V_{eb}^{\rm SR}$ is, accordingly, 
the interaction between the electrons and a rigid,
positive, uniform background of density $n=(4\pi r_s^3/3)^{-1}$
\begin{equation}
V_{eb}^{\rm SR}  =  -n\sum_{i=1}^N\int d\xv \,\frac{\erfc(\mu|\rv_i-\xv|)}
{|\rv_i-\xv|}, 
\label{eq_Veb}
\end{equation}
and $V_{bb}^{\rm SR}$ is the corresponding background-background interaction
\begin{equation}
V_{bb}^{\rm SR}  =  
\frac{n^2}{2}\int d\xv\int d\xv' \,\frac{\erfc(\mu|\xv-\xv'|)}
{|\xv-\xv'|}.
\label{eq_Vbb}
\end{equation}
First we calculate the ground-state energy per electron 
of this model hamiltonian, as a function
of the density parameter $r_s$ and of the parameter $\mu$,
with a diffusion Monte Carlo method (Sec.~\ref{sec_DMC}).
Then we derive the asymptotic behaviors of the correlation energy 
$\ec(r_s,\mu)$ (Sec.~\ref{sec_limits}). On these grounds we finally
(Sec.~\ref{sec_param}) present a convenient analytic parametrization
of the correlation energy, thus following in the footsteps from quantum
simulations of the regular jellium model to the best available LDA
functionals.\cite{CA,VWN,PZ,PW92}

\section{DMC calculation of the ground-state energy}
\label{sec_DMC}

A local density functional for the short-range potential
of Eqs.~(\ref{eq_Vee})-(\ref{eq_Vbb}) should recover the 
Ceperley-Alder\cite{CA}
(CA) correlation energy for $\mu\to 0$. In this Section we outline the
implications of this condition on the technical aspects of our calculation,
which is in all respects a standard application of the diffusion Monte Carlo 
method in the Fixed Node approximation (FN--DMC).\cite{mitas} 

The FN--DMC method gives the energy $E_{FN}$ of the lowest--lying Fermionic 
eigenstate of the Hamiltonian which has the same nodes as the
chosen trial function $\Psi_T$. The error in
$E_{FN}$ is variational, and it vanishes as the nodal structure
of $\Psi_T$ approaches the (unknown) nodal structure of the exact
ground state. The simplest choice for the trial function of a homogeneous
fluid\cite{CA} is the Jastrow--Slater form, $\Psi_T(R)=J(R)D(R)$, where
the symmetric Jastrow factor $J(R)=\exp[-\sum_{i<j}u(r_{ij})]$ describes 
pair correlations, and $D$ is the product of one Slater determinant of 
plane waves (PW) for each spin component ($R$ denotes the coordinates 
of all the particles). A better nodal structure is provided by the
so--called backflow (BF) wave function.\cite{kwon3d}

The method used in Ref.~\onlinecite{CA} is in principle exact: it starts from
the FN solution and then it performs a ``nodal relaxation'', whereby the
energy converges to the exact ground--state result. However, this second
process is accompanied by an increasing statistical noise, which may hinder
full convergence of the results.
In practice, the results of Ref.~\onlinecite{CA} are between the
FN energies recently calculated with PW and BF nodes\cite{kwon3d}, and 
actually somewhat closer to the former. 
Since, on one hand, BF calculations are considerably more demanding, and, on the other, the most widely used local-density functionals are constructed to fit the quantum Monte Carlo results of Ref.~\onlinecite{CA}, we choose to stick to the simple trial
function with Slater determinants of plane waves. In this way our ``short-range local-density functional" will continuously merge into the Ceperley-Alder\cite{CA}-based local-density functionals as $\mu \rightarrow 0$.

\begin{figure}
\includegraphics[width=6.6cm]{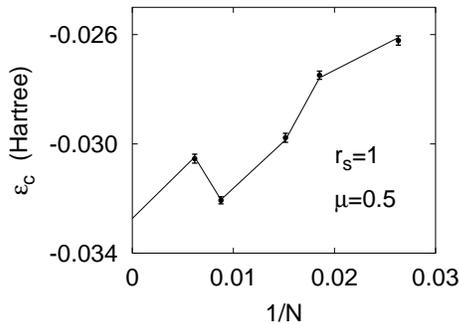} 
\caption{Correlation energy per particle of the short-range interacting gas
at $\mu\!=\!0.5 \,r_s\!=\!1$, for different numbers of particles $N$. 
The fitting
function of Eq.~(\ref{eq_size}) (line) favorably compares with the 
VMC data (dots).
}
\label{fig_size}
\end{figure}

All the other errors in the simulation can be controlled and eliminated.
It is easy to ensure that the biases due to a finite time step and a finite
population of walkers\cite{mitas} are much smaller than the statistical
uncertainty of the CA results, which we set as our target precision.
The number extrapolation is more delicate. We simulate $N$ particles
in a cubic box with periodic boundary conditions, interacting via the
potential of Eq.~(\ref{eq_srpotential}). Since for very small values of $\mu$
we rely on the analytic asymptotic behavior described in Sec.~\ref{sec_limits},
the  only simulations we need to do will deal with really short-range
potentials, which we may safely treat using the minimum image convention.\cite{AT}
The dependence of the energy on the number of particles is determined
with the Variational Monte Carlo (VMC) method, which calculates the 
expectation value of the Hamiltonian operator on the trial wave function
and is cheaper than DMC. 
For several values of $N$ (namely 38, 54, 66, 114, 162), we use VMC
to calculate (i) the variational energy $E_V$ (after optimization of the
Jastrow factor), and (ii) the Hartree--Fock energy $E_{HF}$, which corresponds
to $J=1$. For each value of $r_s$ and $\mu$, the resulting estimate of 
the correlation energy per electron, 
$\epsilon_c=(E_V-E_{HF})/N$, is fitted to the 
following form:
\begin{eqnarray}
\epsilon_c(r_s,\mu;N) & = & \epsilon_c(r_s,\mu;\infty)+a(r_s,\mu)[T(\infty)-T(N)]
\nonumber \\
& & +b(r_s,\mu)/N.
\label{eq_size}
\end{eqnarray}
Here $T(N)$ is the kinetic energy of $N$ non--interacting electrons at $r_s=1$,
and $a(r_s,\mu)$, $b(r_s,\mu)$ and the correlation energy in the
thermodynamic limit, $\epsilon_c(r_s,\mu;\infty)$ are fitting parameters.
The size dependence of the VMC result for the correlation energy is
shown in  Fig.~\ref{fig_size} for the case where it is largest (small
$r_s$  and small $\mu$). We point out that the simple functional guess
of  Eq.~(\ref{eq_size}) (solid line) accurately models the size
dependence of the VMC data which, although
on a small energy scale, are still far from a smooth dependence (dots
with error bars).
Our final result for the correlation energy is obtained by adding the
infinite-size extrapolation obtained from Eq.~(\ref{eq_size}) to the
result of a single DMC simulation with $N=54$.

\section{Asymptotic behaviors}
\label{sec_limits}
In this section we derive some limiting behaviors of the
correlation energy $\ec(r_s,\mu)$, which will be
used for its parametrization in
Sec.~\ref{sec_param}. The detailed study carried out here
can be also of interest for the choice of a density-dependent
$\mu$ parameter in the mixed schemes of 
Refs.~\onlinecite{sav2002} and~\onlinecite{kohn}.
\par
We consider two different regimes: when our system 
approaches the standard jellium model (i.e., full interaction $1/r$),
and when it approaches the noninteracting Fermi gas. 
In the first case (Secs.~\ref{picmu} and~\ref{picrs}) we find that the 
correlation energy
is a function of the scaled variable $\mu \sqrt{r_s}$, while in
 the second case (Sec.~\ref{granmurs}) the relevant scaled variable is $\mu\, r_s$.
\subsection{Finite $r_s$, and $\mu\to 0$}
\label{picmu}
Since for small $\mu$
\begin{equation}
\frac{\erfc(\mu x)}{x}=\frac{1}{x}-\frac{2\mu}{\sqrt{\pi}}+\frac{2}{3}
\frac{x^2}{\sqrt{\pi}}\mu^3+O(\mu^5),
\end{equation}
if we fix the density  and let the parameter $\mu$
approach zero, we can write
\begin{equation}
H=H_{\rm Coul}+\mu H^{(1)}+\mu^3 H^{(3)}+O(\mu^5),
\label{eq_Hpert}
\end{equation}
where 
\begin{eqnarray}
H^{(1)}&=&\frac{N}{\sqrt{\pi}} \\
H^{(3)} & = & \frac{2}{3\sqrt{\pi}}\biggl(\frac{1}{2}\sum_{i\ne j}
|\rv_i-\rv_j|^2-n\sum_i\int d\xv \,|\xv-\rv_i|^2 \nonumber \\
& & +\frac{n^2}{2}
\int d\xv\int d\xv' \,|\xv-\xv'|^2\biggr).
\end{eqnarray}
In Eq.~(\ref{eq_Hpert}), and in the rest of this paper, the suffix ``Coul''
indicates quantities of the standard uniform electron gas (jellium),
with Coulomb interaction $1/r$. 
Thus, for small $\mu$ we are perturbing the jellium model,
\begin{eqnarray}
E(\mu)& =&E_{\rm Coul}+\mu E^{(1)}+\mu^2 E^{(2)}+...
\label{eq_Epert}\\
\Psi(\mu) & = & \Psi_{\rm Coul} +\mu \Psi^{(1)}+\mu^2 \Psi^{(2)}+...
\end{eqnarray}
Since $H^{(1)}$ is a constant, we immediately find $E^{(1)}=H^{(1)}=
N/\sqrt{\pi}$ and $\Psi^{(1)}=0$, which, combined with  $H^{(2)}=0$, also
gives $\Psi^{(2)}=0$ and $E^{(2)}=0$. Then 
$E^{(3)}$ is simply
\begin{equation}
E^{(3)}=\langle \Psi_{\rm Coul}|H^{(3)}|\Psi_{\rm Coul}\rangle,
\label{eq_E3}
\end{equation}
and can be easily 
evaluated, since it is related to the
plasma oscillation,\cite{Pines}
\begin{equation}
E^{(3)}=-\frac{N}{\sqrt{\pi}\,\omega_p}=
-N\frac{r_s^{3/2}}{\sqrt{3\pi}}.
\end{equation}
Eqs.~(\ref{eq_Epert})-(\ref{eq_E3}) hold because
the expectation values of $H^{(1)}$ and $H^{(3)}$ 
on $\Psi_{\rm Coul}$ exist, as it will be more explicitly shown in
Eqs.~(\ref{eq_HeF})-(\ref{eq_prova}).
\par   
 Taking the energy per particle $\epsilon=E/N$, and
dividing it into the non-interacting kinetic part $t_s=\tfrac{3}{10} k_F^2$ 
and the exchange-correlation contribution $\exc$,
we then have the small-$\mu$ expansion
\begin{equation}
\exc(r_s,\mu\to 0)=\exc^{\rm Coul}(r_s)+\frac{\mu}{\sqrt{\pi}}- 
\frac{r_s^{3/2}}{\sqrt{3\pi}}\mu^3 + O(\mu^4).
\label{eq_expexc}
\end{equation}
The same result can be obtained by differentiation of $E(\mu)$ with
respect to $\mu$ and by using the Helmann-Feynmann theorem, which leads
to the exact expression (see also Ref.~\onlinecite{Sav_madrid}):
\begin{equation}
\frac{\partial\exc}{\partial\mu}=-\frac{3}{\sqrt{\pi}}\int_0^{\infty}
ds\,s^2\,e^{-\mu^2r_s^2s^2}[g_{xc}(s,r_s,\mu)-1], 
\label{eq_HeF}
\end{equation}
where $s=r/r_s$, and $g_{xc}(s,r_s,\mu)$ is the 
pair-distribution function\cite{GSB,GP2,PW}
corresponding to the Hamiltonian of Eq.~(\ref{eq_ham}).
The evaluation of
Eq.~(\ref{eq_HeF}) at $\mu=0$, immediately gives the first-order
result, $1/\sqrt{\pi}$.
Higher-order derivatives of $\exc$ at $\mu=0$ can be obtained by further 
differentiating Eq.~(\ref{eq_HeF}), provided that the 
conditions for differentiation under the integral sign are fulfilled.
Since $\Psi^{(1)}=\Psi^{(2)}=0$ implies
$\partial g_{xc}(s,r_s,\mu)/\partial \mu|_{\mu=0}=0$ and
$\partial^2 g_{xc}(s,r_s,\mu)/\partial \mu^2|_{\mu=0}=0$, the possibility
of extracting the second and third derivatives of $\exc$ at $\mu=0$ 
from Eq.~(\ref{eq_HeF}) depends on whether the integrals
\begin{equation}
\int_0^{\infty} ds\,s^n [g_{xc}^{\rm Coul}(s,r_s)-1]
\end{equation}
with $n=4$ and $n=6$ exist. This is the case, since
$g_{xc}^{\rm Coul}(s,r_s)-1$ is a well-behaved function whose
oscillation-averaged part\cite{nota} goes to
zero as\cite{GSB,GP2} $1/s^8$ when $s\to\infty$. We thus obtain
from Eq.~(\ref{eq_HeF})
\begin{eqnarray}
\frac{\partial^2 \exc}{\partial \mu^2}\Big|_{\mu=0}& = & 0 \\
\frac{\partial^3 \exc}{\partial \mu^3}\Big|_{\mu=0}& = &
\frac{6}{\sqrt{\pi}}r_s^2\int_0^{\infty} ds\,s^4 [g_{xc}^{\rm Coul}(s,r_s)-1]
\nonumber \\
& = & \frac{6}{\sqrt{\pi}}r_s^2\left(-\frac{1}{r_s^2\omega_p}\right)=
-\frac{6}{\sqrt{3\pi}}r_s^{3/2}, 
\label{eq_prova}
\end{eqnarray}
in agreement with Eq.~(\ref{eq_expexc}). We see that since
$g_{xc}^{\rm Coul}(s\to\infty,r_s)-1\propto 1/s^8$ no further information
can be extracted from Eq.~(\ref{eq_HeF}), or, equivalently, by going
further with the expansion of Eq.~(\ref{eq_Hpert}).\par

One can divide $\exc$  into its exchange and correlation parts,
$\exc=\ex+\ec$. The exchange energy $\ex$
has been calculated by Savin,\cite{Sav1}
and is reported in Appendix~\ref{app_ex}. Its small-$\mu$
expansion is
\begin{equation}
\ex(r_s,\mu\to 0)=\ex^{\rm Coul}(r_s)+\frac{\mu}{\sqrt{\pi}}
-\frac{3\alpha r_s}{2\pi}
\mu^2+O(\mu^4),
\label{eq_exexp}
\end{equation}
where $\alpha=(4/9\pi)^{1/3}$.
The $\mu\to 0$ behavior of $\ec=\exc-\ex$, is then
\begin{equation}
\ec(r_s,\mu\to0)=\ec^{\rm Coul}(r_s)+\frac{3\alpha r_s}{2\pi}\mu^2-
\frac{r_s^{3/2}}{\sqrt{3\pi}}\mu^3+O(\mu^4).
\label{eq_ecexp}
\end{equation}
Notice that if we divide the pair-distribution function $g_{xc}$ into
its exchange and correlation parts, $g_{xc}(s,r_s,\mu)=g_x(s)+g_c(s,
r_s,\mu)$, we have 
\begin{eqnarray}
\frac{\partial\ex}{\partial\mu}& =& -\frac{3}{\sqrt{\pi}}\int_0^{\infty}
ds\,s^2\,e^{-\mu^2r_s^2s^2}[g_x(s)-1], \label{eq_Hex}\\ 
\frac{\partial\ec}{\partial\mu}& =& -\frac{3}{\sqrt{\pi}}\int_0^{\infty}
ds\,s^2\,e^{-\mu^2r_s^2s^2}g_c(s,r_s,\mu). \label{eq_Hec}
\end{eqnarray}
(This follows directly from the Hellmann-Feynmann theorem 
and from the fact that $g_x$ corresponds to the noninteracting gas and thus
does not depend on $\mu$.)
If we take the limit $\mu\to 0$ of Eqs.~(\ref{eq_Hex}) and~(\ref{eq_Hec}) 
we recover the first-order result in
Eqs.~(\ref{eq_exexp}) and~(\ref{eq_ecexp}). However, higher-order
derivatives at $\mu=0$ of $\ex$ and $\ec$ cannot be obtained
by differentiating Eqs.~(\ref{eq_Hex}) and~(\ref{eq_Hec}). This is due
to the long-range tail of $g_x(s)-1$ and 
$g_c^{\rm Coul}(s,r_s)$: when $s\to \infty$ they both approach 
zero as\cite{PW,GSB,GP2}
$1/s^4$. Thus,
integrals of the kind $\int_0^{\infty}s^4[g_x-1]$ and
$\int_0^{\infty}s^4g_c^{\rm Coul}$ diverge. The long-range tails of
$g_x(s)-1$ and $g_c^{\rm Coul}(s,r_s)$ exactly cancel\cite{PW,GSB,GP2} 
in $g_{xc}^{\rm Coul}(s,r_s)-1$.
This is why, at small $\mu$,  
both $\ex$ and $\ec$ have terms $\propto \mu^2$ which cancel out in $\exc$.
\subsection{Finite $\mu$, and $r_s\to 0$}
\label{picrs}
If we use the relevant  scaled units $\sv= \rv/r_s$ and we let
$r_s$ approach zero, the potential has the expansion
\begin{equation*}
\frac{1}{r_s}\frac{\erfc(\mu r_s s)}{s}\Big|_{r_s\to 0}=
\frac{1}{r_s}\left(\frac{1}{s}-\frac{2\mu}{\sqrt{\pi}}r_s
+\frac{2}{3}\frac{\mu^3 s^2}{\sqrt{\pi}}r_s^3+...\right),
\end{equation*}
which has
the Coulomb interaction as leading term.
We are thus approaching again the jellium model, so that
Eq.~(\ref{eq_ecexp}) is also valid
for finite $\mu$ and $r_s\to 0$. 
\par
In Eq.~(\ref{eq_ecexp}) the relevant scaled
variable is $\mu \sqrt{r_s}$. This can be understood in
the following way. The Coulomb gas presents
screening effects at lenghts $r\gtrsim 1/q_{\rm TF}\propto \sqrt{r_s}$, where
$q_{\rm TF}$ is the Thomas-Fermi wave vector.
Since the $\erfc$ function amounts to some sort of
artificial screening at lenghts $r\gtrsim \mu^{-1}$, 
the Thomas-Fermi screening appears, exactly as in the Coulomb gas,
when $\sqrt{r_s}\ll \mu^{-1}$ .

\subsection{$\mu r_s \gg 1$}
\label{granmurs}
When $\mu\to\infty$, the potential terms
of Eqs.~(\ref{eq_Vee})-(\ref{eq_Vbb}) rapidly vanish
($V^{\rm SR}\sim e^{-\mu^2r^2}$). In this regime we can treat the
whole potential as a perturbation to the non-interacting Fermi gas.\par
The first-order (in the potential) correction to the
non-interacting energy $t_s$ gives $\ex$
of Appendix~\ref{app_ex}. The second-order term can be computed
by standard Rayleigh-Schr\"odinger perturbation theory
\begin{equation}
\e^{(2)}=-\frac{1}{N}\sum_{n\ne 0}
\frac{\langle 0|V^{\rm SR}| n\rangle\langle n |V^{\rm SR}|0\rangle}{E_n-E_0}.
\label{eq_secondorder}
\end{equation}
As in the case of jellium, $\e^{(2)}$ is the sum of a direct term and
of a second-order exchange term,\cite{Pines} which in Fourier space read
\begin{widetext}
\begin{eqnarray}
\e^{(2)}_{\rm dir} & = & -\frac{3}{16\pi^5}\int d\qv
\left(\frac{1-e^{-q^2k_F^2/4\mu^2}}{q^2}\right)^2\int_{|\kv+\qv|>1}
d\kv\int_{|\pv+\qv|>1}d\pv\,\frac{\theta(1-k)\theta(1-p)}{q^2+\qv\cdot
(\kv+\pv)}
\label{eq_ec2dir} \\
\e^{(2)}_{\rm ex} & = & \frac{3}{32\pi^5}\int d\qv\,
\frac{1-e^{-q^2k_F^2/4\mu^2}}{q^2}\int_{|\kv+\qv|>1}
d\kv\int_{|\pv+\qv|>1}d\pv\,\frac{1-e^{-|\qv+\kv+\pv|^2k_F^2/4\mu^2}}
{|\qv+\kv+\pv|^2}\cdot
\frac{\theta(1-k)\theta(1-p)}{q^2+\qv\cdot
(\kv+\pv)}.
\label{eq_ec2ex}
\end{eqnarray}
\end{widetext}
Here all the momenta are expressed in units of $k_F=(\alpha\,r_s)^{-1}$,
and $\theta(x)$ is the Heaviside step function. 
Now, consider the case $\mu r_s \to \infty$ and 
divide the integral over $q$ in Eqs.~(\ref{eq_ec2dir}) 
and~(\ref{eq_ec2ex}) into two parts:
\begin{equation*}
\int_0^\infty dq = \int_0^{q_1}dq+\int_{q_1}^{\infty}dq, \quad\; 
{\rm with}\;1\ll q_1\ll \mu r_s.
\end{equation*}
In the first part, when $q\in [0,q_1]$, we can write
\begin{eqnarray*}
1-e^{-q^2k_F^2/4\mu^2} & \approx & \frac{q^2k_F^2}{4\mu^2}, \\
1-e^{-|\qv+\kv+\pv|^2k_F^2/4\mu^2} & \approx &
\frac{|\qv+\kv+\pv|^2k_F^2}{4\mu^2}
\end{eqnarray*}
(since $q_1 \ll \mu r_s$, and the integrals of
Eqs.~(\ref{eq_ec2dir}) and~(\ref{eq_ec2ex}) are
restricted to the domain $|\kv|\le 1$, $|\pv|\le 1$). 
Equations~(\ref{eq_ec2dir}) and~(\ref{eq_ec2ex}) then reduce to
integrals of the same kind, which can be summed to yield
\begin{eqnarray*}
\e^{(2)}_{|\qv|\le q_1} & = &
\frac{-3}{32\pi^5}\left(\frac{k_F^2}{4\mu^2}\right)^2\int_{|\qv|\le q_1}
d\qv\int_{|\kv+\qv|>1}
d\kv \\ & & 
\times \int_{|\pv+\qv|>1}d\pv\,\frac{\theta(1-k)\theta(1-p)}{q^2+\qv\cdot
(\kv+\pv)},
\end{eqnarray*}
i.e., they give a term which vanishes as $(\mu r_s)^{-4}$.  In the second
part, $q\in [q_1,\infty)$, having chosen $q_1\gg 1$, we can write
\begin{eqnarray*}
& & |\qv+\kv+\pv|^2\approx q^2, \\
& & \int_{|\kv+\qv|>1}
d\kv \int_{|\pv+\qv|>1}d\pv \frac{\theta(1-k)\theta(1-p)}{q^2+\qv\cdot
(\kv+\pv)} \approx \left(\frac{4\pi}{3}\right)^2\frac{1}{q^2}.
\end{eqnarray*}
Equations~(\ref{eq_ec2dir}) and~(\ref{eq_ec2ex}) again reduce to
integrals of the same kind, which can be summed to yield
\begin{equation}
\e^{(2)}_{|\qv|\ge q_1}=-\frac{2}{3\pi^2}\int_{q_1}^{\infty}dq\,
\left(\frac{1-e^{-q^2k_F^2/4\mu^2}}{q^2}\right)^2.
\label{eq_eclargemurs}
\end{equation}
The right-hand side of Eq.~(\ref{eq_eclargemurs}) 
can be evaluated analytically and then expanded
for $\mu r_s \to \infty$. Its leading term is (correctly) independent
of $q_1$ and equals $-\frac{\sqrt{2}-1}{4\sqrt{\pi}}(\mu r_s)^{-3}$.
We thus have
\begin{equation}
\ec(r_s,\mu)\Big|_{\mu r_s \gg 1}=-
\frac{A}{(\mu r_s)^3}+\frac{B}{(\mu r_s)^4}+...
\label{eq_mugrandi}
\end{equation}
with $A\approx \frac{\sqrt{2}-1}{4\sqrt{\pi}}\approx 0.0584$~Hartree.\par

Since the perturbation series expansion whose second-order term corresponds
to Eq.~(\ref{eq_secondorder}) is done with respect to the whole potential 
$V^{\rm SR}$ and not
with respect to the parameter $\mu$, higher-order terms could also contribute
to the value of $A$. For this reason, in our parametrization 
of $\ec(r_s,\zeta)$
$A$ is left as a free parameter, to be optimized with a fit on 
the DMC data. We expect to find a value of $A$ of the same order of the one
estimated with Eq.~(\ref{eq_secondorder}), since the potential
$V^{SR}$ vanishes very rapidly as $\mu\to\infty$, so that the 
higher-order-term contribution to $A$ should be small. 
\begin{figure}
\includegraphics[width=6.6cm]{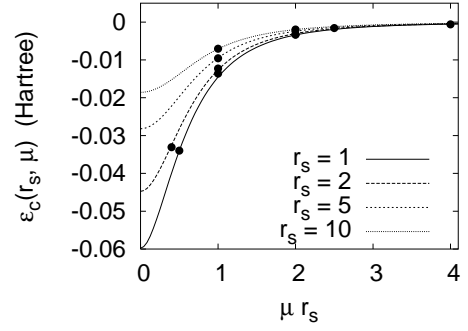} 
\caption{Correlation energy of the short-range interacting gas
as a function of $\mu \,r_s$, for different densities. Our fitting
function (lines) is compared with our DMC data (dots). The error bars
are comparable with the symbol sizes.}
\label{fig_allrs}
\end{figure}
\section{Analytic representation of the correlation energy}
\label{sec_param}
An accurate and simple analytic representation of the correlation
energy $\ec(r_s,\mu)$ can be obtained by a Pad\'e form
which interpolates between the limiting behaviors given
by our Eqs.~(\ref{eq_ecexp}) 
and~(\ref{eq_mugrandi}), and contains some free parameters to fit
our DMC data. We write
\begin{equation}
\ec(r_s,\mu)=\frac{\ec^{\rm Coul}(r_s)[1+b_1(r_s)\mu]}{1+b_1(r_s)\mu
+b_2(r_s)\mu^2+b_3(r_s)\mu^3+b_4(r_s)\mu^4},
\end{equation}
where
\begin{eqnarray}
b_1 & = &\left(b_3-\frac{r_s^{3/2}}{\sqrt{3\pi}}\frac{1}{\ec^{\rm Coul}}
\right)\frac{1}{b_2} \\
b_2 & = & -\frac{3}{2\pi}\frac{\alpha r_s}{\ec^{\rm Coul}} \\
b_4 & = & -b_1\ec^{\rm Coul}\frac{r_s^3}{A},
\end{eqnarray}
and $\ec^{\rm Coul}(r_s)$ is one of the standard 
parametrizations\cite{VWN,PZ,PW92} 
of the correlation energy of the unpolarized jellium. Here
we used the parametrization of Perdew and Wang.\cite{PW92}
The two parameters $b_3(r_s)$ and $A$ are fixed by a two-dimensional 
($r_s,\mu$) best fit to our DMC data. We find:
\begin{eqnarray}
& & b_3(r_s) = 1.27\,r_s^{7/2} \\
& & A = 0.03579. 
\end{eqnarray}
This fit yields a reduced $\chi^2$ of 2.7. In Fig.~\ref{fig_allrs}
we show our DMC data together with the fitting function for different
values of $r_s$. 
Notice that our  analytic $\ec(r_s,\mu)$ does not break down at high 
($r_s\to 0$) or low ($r_s\to\infty$) densities, being constrained by 
exact behaviors.

\section{Conclusions and perspectives}
We have  presented a comprehensive numerical and analytic study of the 
ground-state energy of a (spin unpolarized) uniform electron gas with
modified, short-range-only electron-electron interaction $\erfc(\mu
r)/r$, as  a function of the cutoff parameter $\mu$ and of the electronic 
density. Our chief goal has been the publication, in a convenient form
for  application, of a reliable local density functional for the
correlation energy of this model system, which (i) 
fits the results of our quantum Monte Carlo simulations and (ii)
automatically  incorporates exact limits. Such a functional is a
crucial  ingredient for some recently proposed ``mixed 
schemes",  which exploit the DFT only for the short-range part of the
electron-electron interaction. 
In this context the natural extension of this study will be the generalization
of our functional to the spin-polarized case.
\par
What we obtained in this paper is not the only possible  short-range
local-density functional of interest to ``mixed schemes". In some of
them~\cite{sav2002} the DFT treatment of the short-range part is
handled through another functional 
$\overline{E}_{xc}^{\rm SR}[n]$,
defined as the difference between the standard exchange-correlation
energy functional (corresponding to the Coulomb interaction) and
a long-range-only functional
\begin{equation}
\overline{E}_{xc}^{\rm SR}[n]=E_{xc}[n]-{E}_{xc}^{\rm LR}[n].
\end{equation}
Another direction of future work will thus be the study of the uniform
electron gas with a long-range-only interaction of the form of
Eq.~(\ref{eq_lrpotential}), and, possibly, other modified interactions
proposed in the same spirit.\cite{TSF}

\section*{Acknowledgments}
We thank S. Baroni, A. Savin, and J. Toulouse for 
useful discussions, and gratefully acknowledge financial support from the 
Italian
Ministry of Education, University and Research (MIUR) through COFIN 
2003-2004
and the allocation of computer resources from INFM Iniziativa Calcolo 
Parallelo.
\appendix
\section{Exchange energy}
\label{app_ex}
The exchange energy corresponding to
the Hamiltonian~(\ref{eq_ham}) has been calculated
by Savin in Ref.~\onlinecite{Sav1}, and is equal to
\begin{eqnarray}
\ex(r_s,\mu) & = & -\frac{2}{\pi}k_F\biggl\{\frac{3}{8}-a\Big[\sqrt{\pi}\,
\erf\left(\frac{1}{2a}\right)-3a+4a^3 \nonumber \\
& & +(2a-4a^3)\,e^{-1/4a^2}\Big]\biggr\},
\end{eqnarray}  
with $a=\mu/(2k_F)$. The exchange energy thus satisfies
$$\ex(r_s,\mu)=r_s^{-1}f(\mu r_s).$$



\begin{thebibliography}{99}
\bibitem{nota} Energetically unimportant oscillating terms are dealt with
the evaluation of integrals 
of the kind: $\lim_{a\to 0} \int_0^\infty e^{-a\,x} 
\cos(x)\,dx$.
\end{thebibliography}
\end{document}